\documentclass[12pt]{article}

\usepackage{amsfonts, amsmath}
\newcommand{\be}{\begin{equation}} \newcommand{\ee}{\end{equation}}

\begin{document}
\title{Some Implications of the Density Matrix Deformation in
Statistical Mechanics of the Early Universe}
\thispagestyle{empty}

\author{A.E.Shalyt-Margolin\hspace{1.5mm}\thanks
{Phone (+375) 172 883438; e-mail: a.shalyt@mail.ru;alexm@hep.by}}
\date{}
\maketitle
 \vspace{-25pt}
{\footnotesize\noindent  National Center of Particles and High
Energy Physics, Bogdanovich Str. 153, Minsk 220040, Belarus\\
{\ttfamily{\footnotesize
\\ PACS: 03.65; 05.20
\\
\noindent Keywords: density matrix deformation in statistical
mechanics, average energy deformation, entropy deformation}}

\rm\normalsize \vspace{0.5cm}
\begin{abstract}

This work is an extension of the study into statistical mechanics of
the early Universe that has been the subject in prior works of the
author, the principal approach being the density matrix deformation.
In the work it is demonstrated that the previously derived exponential
ansatz may be successfully applied to the derivation of the free and
average energy deformation as well as entropy deformation. Based on the exponential ansatz, the derivation of a statistical-mechanical
Liouville equation as a deformation of the quantum-mechanical counterpart
is presented. It is shown that deformed Liouville equation will possess
nontrivial components as compared to the normal equation in two cases:
for the original singularity (i.e. early Universe) and for black hole,
that is in complete agreement with the results obtained by the author
with coworkers in earlier works devoted to the deformation in quantum
mechanics at Planck scale. In conclusion some possible applications of
the proposed methods are given, specifically for investigation into
thermodynamics of black holes.

\end{abstract}
\newpage
\section{Introduction}
As is known, both statistical mechanics and quantum mechanics of
the early Universe (Plank scale)are differing from the well-known
analogs \cite{r1},\cite{r2} due to deformation.  The deformation
is understood as a theory extension owing to the introduction of
one or several additional parameters in such a way that the
initial theory appears in the limiting transition \cite{r20}. The
deformation in quantum mechanics at Planck scale takes different
paths: commutator deformation \cite{r3},
\cite{r4},\cite{r5},\cite{r6} or density matrix deformation
\cite{r7}, \cite{r8}. In \cite{r9},\cite{r10} it has been
demonstrated that statistical mechanics of the early Universe,
i.e. at Planck scale, may be constructed with the use of the
above-mentioned density matrix deformation, but now the density
matrix in statistical mechanics.
    In the present work we proceed with a study of deformation in
statistical mechanics of the early Universe.
    In section 2 the attention is directed to the main features of
the Gibbs distribution deformation: using maximum temperature on the
order of the Planck's, that is following from the Generalized
Uncertainty Relations (GUR) in quantum mechanics, we construct
deformation parameter $\tau$; then this parameter is used to introduce
the definition of the density pro-matrix into statistical mechanics and
to derive the primary implications, including the exponential ansatz
giving the statistical density pro-matrix in the explicit form.
         In section 3 the exponential ansatz is used for derivation of the
average energy, free energy and entropy deformations. This ansatz
is used further to study deformation of Liouville equation in
statistical mechanics. It is demonstrated that normal Liouville
equation acquires an additional term in two cases: for the
original singularity and for black hole.
    In conclusion consideration is given to particular applications of the
obtained results, specifically in a study into thermodynamics of black holes.

\section {Deformed Density Matrix in Statistical Mechanics of the
Early Universe}
 In this section we recall the main features for deformation of the
density matrix of the early Universe \cite{r9},\cite{r10}.
 To begin, we consider the Generalized Uncertainty Relations
 "coordinate - momentum" \cite{r4},\cite{r5},\cite{r6}:
\begin{equation}\label{U5}
\triangle x\geq\frac{\hbar}{\triangle p}+\alpha^{\prime}
L_{p}^2\frac{\triangle p}{\hbar}.
\end{equation}
 Using relations (\ref{U5}), it is easy to obtain a similar relation
for the "energy - time" pair. Indeed (\ref{U5}) gives
\begin{equation}\label{U6}
\frac{\Delta x}{c}\geq\frac{\hbar}{\Delta p c
}+\alpha^{\prime}
L_{p}^2\,\frac{\Delta p}{c \hbar},
\end{equation}
then
\begin{equation}\label{U7}
\Delta t\geq\frac{\hbar}{\Delta
E}+\alpha^{\prime}\frac{L_{p}^2}{c^2}\,\frac{\Delta p
c}{\hbar}=\frac{\hbar}{\Delta
E}+\alpha^{\prime}
t_{p}^2\,\frac{\Delta E}{ \hbar}.
\end{equation}
where the smallness of $L_p$ is taken into account so that the
difference between $\Delta E$ and $\Delta (pc)$ can be neglected
and $t_{p}$  is the Planck time
$t_{p}=L_p/c=\sqrt{G\hbar/c^5}\simeq 0,54\;10^{-43}sec$. From
whence it follows that we have a  maximum energy of the order of
Planck's:
\\
$$E_{max}\sim E_{p}$$
\\
Proceeding to the Statistical Mechanics, we further assume that an

internal energy of any ensemble U could not be in excess of
$E_{max}$ and hence temperature $T$ could not be in excess of
$T_{max}=E_{max}/k_{B}\sim T_{p}$. Let us consider density matrix
in Statistical Mechanics (see \cite{r11}, Section 2, Paragraph 3):
\begin{equation}\label{U8}
\rho_{stat}=\sum_{n}\omega_{n}|\varphi_{n}><\varphi_{n}|,
\end{equation}
where the probabilities are given by
\\
$$\omega_{n}=\frac{1}{Q}\exp(-\beta E_{n})$$ and
\\
$$Q=\sum_{n}\exp(-\beta E_{n})$$
\\
Then for a canonical Gibbs ensemble the value
\begin{equation}\label{U9}
\overline{\Delta(1/T)^{2}}=Sp[\rho_{stat}(\frac{1}{T})^{2}]
-Sp^{2}[\rho_{stat}(\frac{1}{T})],
\end{equation}
is always equal to zero, and this follows from the fact that
$Sp[\rho_{stat}]=1$. However, for very high temperatures $T\gg0$
we have $\Delta (1/T)^{2}\approx 1/T^{2}\geq 1/T_{max}^{2}$. Thus,
for $T\gg0$ a statistical density matrix $\rho_{stat}$ should be
deformed so that in the general case
\begin{equation}\label{U10}
Sp[\rho_{stat}(\frac{1}{T})^{2}]-Sp^{2}[\rho_{stat}(\frac{1}{T})]
\approx \frac{1}{T_{max}^{2}},
\end{equation}
or \begin{equation}\label{U11}
Sp[\rho_{stat}]-Sp^{2}[\rho_{stat}] \approx
\frac{T^{2}}{T_{max}^{2}},
\end{equation}
In this way $\rho_{stat}$ at very high $T\gg 0$ becomes dependent
on the parameter $\tau = T^{2}/T_{max}^{2}$, i.e. in the most
general case
\\
$$\rho_{stat}=\rho_{stat}(\tau)$$ and $$Sp[\rho_{stat}(\tau)]<1$$
\\
and for $\tau\ll 1$ we have $\rho_{stat}(\tau)\approx\rho_{stat}$
(formula (\ref{U8})) .\\ This situation is identical to the case
associated with the deformation parameter $\alpha = l_{min}^{2
}/x^{2}$ of QMFL given in section â 2 \cite{r8}. That is the
condition $Sp[\rho_{stat}(\tau)]<1$ has an apparent physical
meaning when:
\begin{enumerate}
 \item At temperatures close to $T_{max}$ some portion of information
about the ensemble is inaccessible in accordance with the
probability that is less than unity, i.e. incomplete probability.
 \item And vice versa, the longer is the distance from $T_{max}$ (i.e.
when approximating the usual temperatures), the greater is the
bulk of information and the closer is the complete probability to
unity.
\end{enumerate}
 Therefore similar to the introduction of the deformed

quantum-mechanics density matrix in section 3 \cite{r8},we give
the following
\\
\noindent {\bf Definition.} {\bf(Deformation of Statistical
Mechanics)} \noindent \\Deformation of Gibbs distribution valid
for temperatures on the order of the Planck's $T_{p}$ is described
 by deformation of a statistical density matrix
  (statistical density pro-matrix) of the form
\\$${\bf \rho_{stat}(\tau)=\sum_{n}\omega_{n}(\tau)|\varphi_{n}><\varphi_{n}|}$$
 having the deformation parameter
$\tau = T^{2}/T_{max}^{2}$, where
\begin{enumerate}
\item $0<\tau \leq 1/4$;
\item The vectors $|\varphi_{n}>$ form a full orthonormal system;
\item $\omega_{n}(\tau)\geq 0$ and for all $n$ at $\tau \ll 1$
 we obtain
 $\omega_{n}(\tau)\approx \omega_{n}=\frac{1}{Q}\exp(-\beta E_{n})$
In particular, $\lim\limits_{T_{max}\rightarrow \infty
(\tau\rightarrow 0)}\omega_{n}(\tau)=\omega_{n}$
\item
$Sp[\rho_{stat}]=\sum_{n}\omega_{n}(\tau)<1$,
$\sum_{n}\omega_{n}=1$;
\item For every operator $B$ and any $\tau$ there is a
mean operator $B$ depending on  $\tau$ \\
$$<B>_{\tau}=\sum_{n}\omega_{n}(\tau)<n|B|n>.$$
\end{enumerate}
Finally, in order that our Definition  agree with the formula
(\ref{U11}), the following condition must be fulfilled:
\begin{equation}\label{U12}
Sp[\rho_{stat}(\tau)]-Sp^{2}[\rho_{stat}(\tau)]\approx \tau.
\end{equation}
Hence we can find the value for $Sp[\rho_{stat}(\tau)]$
 satisfying
the condition of Definition 2 (similar to Definition 1):
\begin{equation}\label{U13}
Sp[\rho_{stat}(\tau)]\approx\frac{1}{2}+\sqrt{\frac{1}{4}-\tau}.
\end{equation}
It should be noted:

\begin{enumerate}
\item The condition $\tau \ll 1$ means that $T\ll T_{max}$ either
$T_{max}=\infty$ or both in accordance with a normal Statistical
Mechanics and canonical Gibbs distribution (\ref{U8})
\item Similar to QMFL in \cite{r7},\cite{r8}, where the deformation
parameter $\alpha$ should assume the value $0<\alpha\leq1/4$. As
seen from (\ref{U13}), here $Sp[\rho_{stat}(\tau)]$ is well
defined only for $0<\tau\leq1/4$. This means that the feature
occurring in QMFL at the point of the fundamental length
$x=l_{min}$ in the case under consideration is associated with the
fact that {\bf highest  measurable temperature of the ensemble is
always} ${\bf T\leq \frac{1}{2}T_{max}}$.

\item The constructed deformation contains all four fundamental constants:
 $G,\hbar,c,k_{B}$ as $T_{max}=\varsigma T_{p}$,where $\varsigma$
 is the denumerable function of  $\alpha^{\prime}$
(\ref{U5})and $T_{p}$, in its turn, contains all the above-mentioned
 constants.

\item Again similar to QMFL, as a possible solution for (\ref{U12})
we have an exponential ansatz
\\
$$\rho_{stat}^{*}(\tau)=\sum_{n}\omega_{n}(\tau)|n><n|=\sum_{n}
exp(-\tau) \omega_{n}|n><n|$$
\\
\begin{equation}\label{U14}
Sp[\rho_{stat}^{*}(\tau)]-Sp^{2}[\rho_{stat}^{*}(\tau)]=\tau+O(\tau^{2}).
\end{equation}
In such a way with the use of an exponential ansatz (\ref{U14})
the deformation of a canonical Gibbs distribution at Planck scale
(up to factor $1/Q$) takes an elegant and completed form:
\begin{equation}\label{U15}
{\bf \omega_{n}(\tau)=exp(-\tau)\omega_{n}= exp(-\frac{T^{2}}
{T_{max}^{2}}-\beta E_{n})}
\end{equation}
where $T_{max}= \varsigma T_{p}$

\section{Some Implications}
Using in this section only the exponential ansatz of (\ref{U14}),
in the coordinate representation we have the following:
\begin{equation}\label{U16}
\rho(x,x^{\prime},\tau)=\sum_{i}\frac{1}{Q}e^{-\beta
E_{i}-\tau}\varphi_{i}(x)\varphi_{i}^{*}(x^{\prime})
\end{equation}
However, as $H \mid \varphi_{i}>=E_{i} \mid \varphi_{i}>$, then
\begin{equation}\label{U17}
\rho(\beta,\tau)=\frac{1}{Q}\sum_{i}e^{-\beta H-\tau}\mid
\varphi_{i}><\varphi_{i}\mid=\frac{e^{-\beta H-\tau}}{Q},
\end{equation}
where $Q=\sum_{i}e^{-\beta E_{i}}=Spe^{-\beta H}$. Consequently,
\begin{equation}\label{U18}
\rho(\beta,\tau)=\frac{e^{-\beta H-\tau}}{Spe^{-\beta H}}
\end{equation}
In this way the "deformed" average energy of a system is obtained as
\begin{equation}\label{U19}
U_{\tau}=Sp\rho(\tau)H=\frac{He^{-\beta H-\tau}}{Spe^{-\beta H}}
\end{equation}
The calculation of "deformed" entropy is also a simple task.
Indeed, in the general case of the canonical Gibbs distribution
the probabilities are equal
\begin{equation}\label{U20}
P_{n}=\frac{1}{Q}e^{-\beta E_{n}}
\end{equation}
Nevertheless, in case under consideration they are "replenished" by
$exp(-\tau)$ factor and hence are equal to
\begin{equation}\label{U21}
P^{\tau}_{n}=\frac{1}{Q}e^{-(\tau+\beta E_{n})}
\end{equation}
Thus, a new formula for entropy in this case is as follows:
\begin{equation}\label{U22}
S_{\tau}=-k_{B}e^{-\tau}\sum_{n}P_{n}(lnP_{n}-\tau)
\end{equation}
It is obvious that
 $\lim\limits_{\tau\rightarrow 0}S_{\tau}=
S$, where $S$ - entropy of the canonical ensemble, that is a complete
analog of its counterpart in quantum mechanics at Planck scale
$\lim\limits_{\alpha\rightarrow 0}S_{\alpha}= S$,
where $S$ - statistical entropy in quantum mechanics, and deformation
parameter $\tau$ is changed by $\alpha$ \cite{r8},\cite{r10}.
\\Given the average energy deformation in a system $U_{\tau}$ and
knowing the entropy deformation, one is enabled to calculate the
"deformed" free energy $F_{\tau}$ as well:
\begin{equation}\label{U23}
F_{\tau}=U_{\tau}-TS_{\tau}
\end{equation}
Consider the counterpart of Liouville equation
\cite{r11} for the unnormed
$\rho(\beta,\tau)$ (\ref{U18}):
\begin{equation}\label{U24}
-\frac{\partial\rho(\beta,\tau)}{\partial\beta}=
-\frac{\partial}{\partial\beta}e^{-\tau-\beta H}
\end{equation}
Since
\\
$$\tau=\frac{T^{2}}{T_{max}^{2}}=\frac{\beta^{2}_{max}}{\beta^{2}},$$
\\
where $\beta_{max}=1/k_{B}T_{max}\sim 1/k_{B}T_{P}\equiv \beta_{P}$,
$\tau=\tau(\beta)$. Taking this into consideration and expanding
the right-hand side of equation (\ref{U24}), we get deformation  of
Liouville equation further referred to as $\tau$-deformation:
\begin{equation}\label{U25}
-\frac{\partial\rho(\beta,\tau)}{\partial\beta}=
-e^{-\tau}\frac{\partial \tau}{\partial \beta}+e^{-\tau}H
\rho(\beta)=e^{-\tau}[H \rho(\beta)-\frac{\partial \tau}{\partial
\beta}],
\end{equation}
where $\rho(\beta)=\rho(\beta,\tau=0)$.
\\ The first term in brackets (\ref{U25}) generates Liouville equation.
Actually, taking the limit of the left and right sides
(\ref{U25}) for $\tau\rightarrow 0$, we derive the normal Liouville
equation for $\rho(\beta)$ in statistical mechanics \cite{r11}:
\begin{equation}\label{U26}
-\frac{\partial\rho(\beta)}{\partial\beta}=H \rho(\beta)
\end{equation}
By this means we obtain a complete analog of the
quantum-mechanical results for the associated deformation of
Liouville equation derived  in \cite{r7},\cite{r8},\cite{r12}.
Namely:
\\
(1)early Universe (scales approximating those of the Planck's,
original singularity, $\tau>0$). The density pro-matrix $\rho(\beta,\tau)$
is introduced and a $\tau$-"deformed" Liouville equation(\ref{U25}),
respectively;
\\
(2)after the inflation extension (well-known scales, $\tau\approx
0$) the normal density matrix $\rho(\beta)$ appears in the limit
$\lim\limits_{\tau\rightarrow 0}\rho(\beta,\tau)=\rho(\beta)$.
$\tau$-"deformation" of Liouville equation (\ref{U25}) is changed
by a well-known Liouville equation(\ref{U26});
\\
(3)and finally the case of the matter absorbed by a black hole and
its tendency to the singularity. Close to the black hole
singularity both quantum and statistical mechanics are subjected
to deformation as they do in case of the original singularity
\cite{r7}--\cite{r9},\cite{r12},\cite{r13}. Introduction of
temperature on the order of the Planck's \cite{r14},\cite{r15} and
hence the deformation parameter $\tau > 0$ may be taken as an
indirect evidence for the fact. Because of this, the case is
associated with the reverse transition from the well-known density
matrix in statistical mechanics $\rho(\beta)$ to its
$\tau$-"deformation" $\rho(\beta,\tau)$ and from Liouville
equation(\ref{U26}) to its $\tau$-"deformation" (\ref{U25}.

\section{Conclusion}
This paper is an extension of the study presented in
\cite{r9},\cite{r10}. There is a reason to believe that the
approaches proposed by the author and notions of the "average
energy deformation", "free energy deformation", "entropy
deformation" and so on may be applied during investigation into
thermodynamics of black holes too \cite{r16}. The problem of their
relation to the methods used in studies of black holes with due
regard for GUR \cite{r17}--\cite{r19} is of particular importance.
Of interest is also elucidation of the following facts: are there
any physically meaningful deformations following from the
solutions for(\ref{U12})which differ from the exponential ansatz,
or is it unique in a sense? All these problems require further
investigation.
\end{enumerate}

%References

\end{document}